\documentclass[12pt]{article}
\usepackage{graphicx}
\usepackage{color}
\def\hybrid{\topmargin 0pt      \oddsidemargin 0pt
        \headheight 0pt \headsep 0pt
       \voffset-1cm
        \textwidth 6.25in       % A4 paper
       \textheight 9.5in       % A4 paper
        \marginparwidth 0.0in
        \parskip 5pt plus 1pt   \jot = 1.5ex}
\catcode`\@=11
\def\marginnote#1{}

\newcount\hour
\newcount\minute
\newtoks\amorpm
\hour=\time\divide\hour by60
\minute=\time{\multiply\hour by60 \global\advance\minute by-\hour}
\edef\standardtime{{\ifnum\hour<12 \global\amorpm={am}%
        \else\global\amorpm={pm}\advance\hour by-12 \fi
        \ifnum\hour=0 \hour=12 \fi
        \number\hour:\ifnum\minute<10 0\fi\number\minute\the\amorpm}}
\edef\militarytime{\number\hour:\ifnum\minute<10 0\fi\number\minute}

\def\draftlabel#1{{\@bsphack\if@filesw {\let\thepage\relax
   \xdef\@gtempa{\write\@auxout{\string
      \newlabel{#1}{{\@currentlabel}{\thepage}}}}}\@gtempa
   \if@nobreak \ifvmode\nobreak\fi\fi\fi\@esphack}
        \gdef\@eqnlabel{#1}}
\def\@eqnlabel{}
\def\@vacuum{}
\def\draftmarginnote#1{\marginpar{\raggedright\scriptsize\tt#1}}

\def\draftlabel#1{{\@bsphack\if@filesw {\let\thepage\relax
   \xdef\@gtempa{\write\@auxout{\string
      \newlabel{#1}{{\@currentlabel}{\thepage}}}}}\@gtempa
   \if@nobreak \ifvmode\nobreak\fi\fi\fi\@esphack}
        \gdef\@eqnlabel{#1}}
\def\@eqnlabel{}
\def\@vacuum{}
\def\draftmarginnote#1{\marginpar{\raggedright\scriptsize\tt#1}}

\def\draft{\oddsidemargin -.5truein
        \def\@oddfoot{\sl preliminary draft \hfil
        \rm\thepage\hfil\sl\today\quad\militarytime}
        \let\@evenfoot\@oddfoot \overfullrule 3pt
        \let\label=\draftlabel
        \let\marginnote=\draftmarginnote
   \def\@eqnnum{(\theequation)\rlap{\kern\marginparsep\tt\@eqnlabel}%
\global\let\@eqnlabel\@vacuum}  }

\def\numberbysection{\@addtoreset{equation}{section}
        \def\theequation{\thesection.\arabic{equation}}}

\def\underline#1{\relax\ifmmode\@@underline#1\else
        $\@@underline{\hbox{#1}}$\relax\fi}

\def\titlepage{\@restonecolfalse\if@twocolumn\@restonecoltrue\onecolumn
     \else \newpage \fi \thispagestyle{empty}\c@page\z@
        \def\thefootnote{\fnsymbol{footnote}} }

\def\endtitlepage{\if@restonecol\twocolumn \else  \fi
        \def\thefootnote{\arabic{footnote}}
        \setcounter{footnote}{0}}  %\c@footnote\z@ }
\relax

\hybrid

\newfont{\Bbb}{msbm10 scaled 1\@ptsize00}
\newfont{\Bbbb}{msbm7 scaled 1\@ptsize00}
\newcommand{\CC}{\mbox{\Bbb C}}

\newcommand{\DDD}{\raise-1pt\hbox{$\mbox{\Bbbb D}$}}

\newcommand{\UUU}{\raise-1pt\hbox{$\mbox{\Bbbb U}$}}

\newcommand{\z}{\raise-1pt\hbox{$\mbox{\Bbbb Z}$}}

\def\beq{\begin{equation}}
\def\eeq{\end{equation}}
\def\p{\partial}

\begin{document}

\begin{titlepage}

\title{Asymmetric 6-vertex model and classical Ruijsenaars-Schneider system
of particles}

\author{A. Liashyk\thanks{
BITP, Metrolohichna str. 14-b, 03680, Kiev, Ukraine;
National Research University Higher School of Economics, Russian Federation;
Skolkovo Institute of Science and Technology, 143026 Moscow, Russian Federation,
e-mail:  a.liashyk@gmail.com }
\and D.~Rudneva\thanks{National Research University Higher School of 
Economics,  
Russian Federation;
University of California Santa Barbara, Santa Barbara, CA 93106, USA;
e-mail: missdaryarudneva@gmail.com}
\and A.~Zabrodin
\thanks{ 
National Research University Higher School of Economics,
Russian Federation; 
ITEP, 25
B.Cheremushkinskaya, Moscow 117218, Russia;
Skolkovo Institute of Science and Technology, 143026 Moscow, Russian Federation;
e-mail: zabrodin@itep.ru}
\and 
A. Zotov\thanks{Steklov Mathematical
Institute of Russian Academy of Sciences, Gubkina str. 8, 119991,
Moscow, Russia; ITEP, 25 B.Cheremushkinskaya, Moscow 117218, Russia;
MIPT, Inststitutskii per.  9, Dolgoprudny, Moscow region, 141700,
Russia; e-mail: zotov@mi.ras.ru} }

\date{November 2016}
\maketitle

\vspace{-7cm} \centerline{ \hfill ITEP-TH-27/16}\vspace{7cm}

\begin{abstract}

We discuss the correspondence between models solved by Bethe ansatz
and classical integrable systems of Calogero type.
We illustrate the correspondence by the simplest example of the 
inhomogeneous asymmetric 6-vertex model
parametrized by trigonometric (hyperbolic) functions.

\end{abstract}

\end{titlepage}

\vspace{5mm}

\section{Introduction}

The correspondence between quantum or statistical models solved by Bethe ansatz
and classical integrable many-body systems of Calogero type (the quantum-classical 
duality) was established in \cite{AKLTZ13} for the case of XXX type models. 
See also \cite{ALTZ14,GZZ14,Z2,TZZ15,MTV12,NRS11} for different aspects of this
remarkable correspondence.
It was extended to models of the XXZ type related 
to quantum affine algebras $U_q(\widehat{sl_N})$ in 
\cite{Z1,BLZZ16}.
In this paper we illustrate the correspondence by the simplest example of the 
inhomogeneous asymmetric 6-vertex model
parametrized by trigonometric (hyperbolic) functions which is related to 
the trigonometric (hyperbolic) Ruijsenaars-Schneider system of particles \cite{Ruijs1}.

The asymmetric 6-vertex model can be thought of as the symmetric one 
in horizontal and vertical external fields \cite{Baxter,R}.
In a natural basis,
the matrix of Boltzmann weights of the asymmetric 6-vertex model has the form
$$
{\sf R}=\left ( \begin{array}{cccc}
a&0&0&0\\ 0&b&c&0 \\
0&c'&b'&0\\ 0&0&0&a'\end{array}\right ).
$$
The standard argument shows that the partition function 
with periodic boundary condition depends only on the product
$cc'$, so we can put $c'=c$ from the very beginning without loss of generality.

In this paper we consider the integrable inhomogeneous  
6-vertex model on $L$ sites, with $x_i$ ($i=1, \ldots , L$) being inhomogeneity parameters 
at the sites of the lattice.
As is known, the asymmetric model with the horizontal external field $h$ 
($a/a'=b/b'=e^{2h}$) is equivalent to
the symmetric one ($a'=a$, $b'=b$) with twisted boundary conditions which preserve integrability. The twist matrix is ${\bf g}=\mbox{diag} \, (e^{Lh}, e^{-Lh})$.
In fact the transfer matrices 
of the two models differ by a similarity transformation, so they have the same spectrum.
The transfer matrix of the twisted model, ${\bf T}^{(h)}(x)$, as 
a function of the spectral parameter $x$, 
has simple poles at the 
points $x_i$. The residues at the poles,
${\bf H}_i=(\sinh \eta )^{-1}\mbox{res}_{x=x_i}{\bf T}^{(h)}(x_i)$, 
where $\eta$ is the anisotropy parameter, are commuting operators 
which can be simultaneously diagonalized. In the framework 
of the quantum-classical duality \cite{GZZ14}, their eigenvalues are to be identified with 
velocities, $\dot x_i$, of the classical Ruijsenaars-Schneider particles while
the inhomogeneity parameters are identified with their coordinates,
$x_i$, with the condition that the higher integrals of motion of the classical 
model take some prescribed values expressed through spectral invariants
of the twist matrix. We will also show that 
eigenvalues of another distinguished set
of commuting operators, ${\bf G}_i={\bf T}^{(h)}(x_i -\eta )$, should be 
identified with $-\eta^{-1}e^{-\eta p_i}$, where $p_i$ are momenta of the 
Ruijsenaars-Schneider particles.

We thus see that the different eigenstates of the transfer matrix
correspond to intersection points of two Lagrangian submanifolds 
in the Ruijsenaars-Schneider phase space: one of them is the hyperplane 
$x_i=\mbox{const}$ and the other one is the level set of classical Hamiltonians 
in involution.

\section{The asymmetric inhomogeneous 6-vertex model}

\paragraph{The symmetric model.}
We start with the well known symmetric 6-vertex model. The  
Boltzmann weights $a'=a$, $b'=b$, $c'=c$ are given by the $R$-matrix
%\beq\label{6v1}
%{\sf R}(x)=\left (
%\begin{array}{cccc}
%\sinh (x+\eta )& 0& 0& 0\\ &&&\\
%0& \sinh x & \sinh \eta & 0 \\ &&& \\
%0& \sinh \eta & \sinh x & 0 \\ &&& \\
%0& 0& 0 &\sinh (x+\eta )\end{array}
%\right )
%\eeq
\beq\label{6v1}
{\sf R}(x)=\left (
\begin{array}{cccc}
\displaystyle{\frac{\sinh (x\! + \! \eta )}{\sinh x}}& 0& 0& 0\\ &&&\\
0& 1& \displaystyle{\frac{\sinh \eta }{\sinh x}} & 0 \\ &&& \\
0&\displaystyle{\frac{\sinh \eta }{\sinh x}} & 1 & 0 \\ &&& \\
0& 0& 0 & \displaystyle{\frac{\sinh (x\! +\! \eta )}{\sinh x}} \end{array}
\right )
\eeq
in the standard trigonometric (hyperbolic) parametrization \cite{Baxter}, 
where $x$ is the spectral parameter and $\eta$ is the anisotropy parameter.

Let $V_i\cong \CC ^2$ be several copies of 
the linear space $\CC ^2$, then ${\sf R}_{ij}(x)$ is a linear operator on 
$\bigotimes_l V_l$ which acts non-trivially in 
$V_i\otimes V_j$. 
This $R$-matrix satisfies the Yang-Baxter equation
\beq\label{yb}
{\sf R}_{12}(x-x'){\sf R}_{13}(x){\sf R}_{23}(x')=
{\sf R}_{23}(x'){\sf R}_{13}(x){\sf R}_{12}(x-x'),
\eeq
where the both sides are operators in $V_1\otimes V_2 \otimes V_3$.
The anisotropy parameters are the same for the three $R$-matrices.

Another important property of the $R$-matrix (\ref{6v1}) is its invariance 
under the diagonal 
Cartan subgroup of
$GL(2)\! \times \! GL(2)$  
which means commutativity 
with ${\bf g}\otimes {\bf g}$, where 
%$\mathbf{g}=\mathrm{diag} (g_1, g_2)$ 
$\mathbf{g}=\left ( \begin{array}{cc}g_1&0\\ 0&g_2 \end{array}\right )$
is any diagonal 2$\times$2 matrix. Set $\mathbf{g}_1 =\mathbf{g}\otimes \mathbf{1}$,
$\mathbf{g}_2 =\mathbf{1} \otimes \mathbf{g}$, then it is easy to check that
\beq\label{6v2}
\mathbf{g}_1 \, \mathbf{g}_2 \, {\sf R}_{12}(x)={\sf R}_{12}(x)\,
\mathbf{g}_1 \, \mathbf{g}_2.
\eeq

\paragraph{The asymmetric model.}
Let us label by $0$ the horizontal (``auxiliary'') space and by 
$i=1, \ldots , L$ the vertical (``quantum'') spaces. The matrices of the Boltzmann 
weights for the asymmetric model with the horizontal field $h$ and the 
vertical field $v$ are defined as 
\beq\label{6v3}
\begin{array}{c}
{\sf R}_{0i}^{h,v} (x)\, =\, e^{\frac{1}{2}\, h \sigma_0^z}
e^{\frac{1}{2}\, v \sigma_i^z}{\sf R}_{0i}(x)\,
e^{\frac{1}{2}\, h \sigma_0^z}
e^{\frac{1}{2}\, v \sigma_i^z}
\\  \\
=\, \left (\begin{array}{cc}e^{h/2}&0\\ 0& e^{-h/2}\end{array}\right )_0
\left (\begin{array}{cc}e^{v/2}&0\\ 0& e^{-v/2}\end{array}\right )_i
{\sf R}_{0i}(x)
\left (\begin{array}{cc}e^{h/2}&0\\ 0& e^{-h/2}\end{array}\right )_0
\left (\begin{array}{cc}e^{v/2}&0\\ 0& e^{-v/2}\end{array}\right )_i\,.
\end{array}
\eeq
The explicit form of this ``asymmetric'' $R$-matrix in the trigonometric 
parametrization is
\beq\label{6v1as}
{\sf R}^{h,v}(x)=\left (
\begin{array}{cccc}
\displaystyle{e^{h+v}\frac{\sinh (x\! + \! \eta )}{\sinh x}}& 0& 0& 0\\ &&&\\
0& e^{h-v}& \displaystyle{\frac{\sinh \eta }{\sinh x}} & 0 \\ &&& \\
0&\displaystyle{\frac{\sinh \eta }{\sinh x}} &e^{-h+v} & 0 \\ &&& \\
0& 0& 0 & e^{-h-v}\displaystyle{\frac{\sinh (x\! +\! \eta )}{\sinh x}} \end{array}
\right ).
\eeq

The Yang-Baxter equation (\ref{yb}) combined with the 
Cartan subgroup invariance property
%$gl(1)\! \times \! gl(1)$-invariance
(\ref{6v2}) implies the following Yang-Baxter equation for the asymmetric $R$-matrices
with the same parameter $\eta$:
\beq\label{yb2}
{\sf R}_{12}^{-v', \, v}(x-x')\, {\sf R}_{13}^{h,v} (x)\, {\sf R}_{23}^{h, v'}\! (x')=
{\sf R}_{23}^{h, v'}\! (x')\, {\sf R}_{13}^{h,v}(x)\, {\sf R}_{12}^{-v',\, v}(x-x').
\eeq
In \cite{BBF}, the existence of an $R$-matrix that intertwines 
${\sf R}^{h,v} (x) $ and ${\sf R}^{h,v'} \! (x)$ was proved by a direct solution 
of the Yang-Baxter equation.

Below we consider the inhomogeneous 
asymmetric 6-vertex model with periodic
boun\-dary condition in the horizontal direction. 
The transfer matrix of the model is defined in the usual way \cite{QISM2}:
\beq\label{6v4}
{\bf T}^{h,v}(x)=\mathrm{tr}_0 \Bigl ( {\sf R}_{01}^{h,v}(x-x_1)\,
{\sf R}_{02}^{h,v}(x-x_2)\, \ldots \, {\sf R}_{0L}^{h,v}(x-x_L)\Bigr ).
\eeq
The inhomogeneity parameters $x_i$ 
at the sites are shifts of the spectral parameter. We assume that they are in general position,
i.e. $x_i\neq x_j$ and $x_i\neq x_j\pm \eta$ for any $i\neq j$.
It follows from the Yang-Baxter equation (\ref{yb2}) that the transfer matrices 
with different $x$'s and $v$'s (but the same $\eta$, $h$ and $\{x_i\}_L$) commute:
$[{\bf T}^{h,v}(x), \, {\bf T}^{h,v'}(x')]=0$.
It is easy to see that 
the dependence of the transfer matrix on the vertical field $v$ is very simple:
\beq\label{6v5}
{\bf T}^{h,v}(x)=e^{v{\bf S}^z}{\bf T}^{h,0}(x),
\eeq
where 
\beq\label{6v6}
{\bf S}^z=\sum_{i=1}^L \sigma_i^z =\mathbf{M}_1-\mathbf{M}_2
\eeq
is the operator that counts the (conserved) difference between the total 
number of up (${\bf M}_1=\frac{1}{2}\sum_{i=1}^L(1+\sigma_i^z)$) and down 
(${\bf M}_2=\frac{1}{2}\sum_{i=1}^L(1-\sigma_i^z)$)
looking
arrows on vertical edges. Note that ${\bf M}_1+{\bf M}_2=L{\bf 1}$, where
${\bf 1}$ is the unity operator.

In fact the transfer matrix of the asymmetric model with periodic boundary 
condition is connected with the transfer matrix 
of the symmetric model with {\it twisted} boundary condition 
by a similarity transformation. Set (cf. \cite{Rittenberg})
$$
\mathbf{U}=\mathbf{1}\otimes e^{h\sigma^z}\otimes e^{2h\sigma^z}\otimes \ldots 
\otimes e^{(L-1)h\sigma^z}=
\exp \left ( \sum_{j=1}^L (j\! -\! 1)h\sigma^z_j\right ).
$$
The Cartan subgroup invariance
(\ref{6v2}) implies the relation
$$
\mathbf{U}{\bf T}^{h,v}(x)\mathbf{U}^{-1}=e^{v{\bf S}^z}
\mathbf{T}^{(h)}(x),
$$
where
\beq\label{6v7}
\mathbf{T}^{(h)}(x)=\mathrm{tr}_0 \Bigl ( {\sf R}_{01}(x-x_1)\,
{\sf R}_{02}(x-x_2)\, \ldots \, {\sf R}_{0L}(x-x_L)e^{Lh\sigma_0^z}\Bigr )
\eeq
is the transfer matrix for the symmetric model with the boundary condition
twisted by the diagonal group element ${\bf g}=e^{Lh\sigma^z}$.

Diagonal matrix elements of the $R$-matrix (\ref{6v1}) are periodic under the 
shift $x\to x+i\pi$ while the non-diagonal ones are anti-periodic. Therefore, 
the trace (\ref{6v7}) enjoys the periodicity condition 
$\mathbf{T}^{(h)}(x+i\pi ) =\mathbf{T}^{(h)}(x)$. Since it has first order poles at the 
points $x_i$, its pole expansion can be written as
\beq\label{6v8}
\mathbf{T}^{(h)}(x)={\bf C}+\sinh \eta 
\sum_{k=1}^L \mathbf{H}_k \coth (x-x_k),
\eeq
where 
$$
\mathbf{C}=\frac{1}{2}\Bigl ( \mathbf{T}^{(h)}(\infty )+\mathbf{T}^{(h)}(-\infty )\Bigr ),
\quad
%\mathbf{H}_k =\frac{\mathrm{res}_{z=x_k}\mathbf{T}^{(h)}(x)}{\sinh \eta}
\mathbf{H}_k =(\sinh \eta )^{-1}\, \mathrm{res}_{z=x_k}\mathbf{T}^{(h)}(z)
$$
are some commuting operators. They can be regarded
as Hamiltonians of an integrable quantum spin chain with long range interaction.
The limiting values of $\mathbf{T}^{(h)}(x)$ as $x\to \pm \infty$ can be easily found:
$$
\begin{array}{l}
\displaystyle{\mathbf{T}^{(h)}(\infty )\, = \, \mathbf{C}+
\sinh \eta \sum_k \mathbf{H}_k \, =\, 
e^{Lh}e^{\eta \mathbf{M}_1}+e^{-Lh}e^{\eta \mathbf{M}_2}},
\\ \\
\displaystyle{\mathbf{T}^{(h)}(-\infty ) = \mathbf{C}-\sinh \eta \sum_k \mathbf{H}_k =
e^{Lh}e^{-\eta \mathbf{M}_1}+e^{-Lh}e^{-\eta \mathbf{M}_2}}.
\end{array}
$$
Therefore, we have the following sum rules:
\beq\label{6v9}
\mathbf{C}=e^{Lh}\cosh (\eta \mathbf{M}_1)+e^{-Lh}\cosh (\eta \mathbf{M}_2),
\eeq
\beq\label{6v10}
\sum_{k=1}^L\mathbf{H}_k= e^{Lh}\, \frac{\sinh (\eta \mathbf{M}_1)}{\sinh \eta}+
e^{-Lh}\, \frac{\sinh (\eta \mathbf{M}_2)}{\sinh \eta}.
\eeq

For the needs of finding the partition function one is interested 
in the solution of the common spectral problem
\beq\label{6v11}
\left \{\begin{array}{l}
\mathbf{T}^{(h)}(x)\Bigl |\Psi \Bigr >=T(x)\Bigl |\Psi \Bigr >
\\ \mathbf{M}_1\Bigl |\Psi \Bigr >=M_1\Bigl |\Psi \Bigr >
\end{array}\right. \quad \mbox{or} \quad
\left \{\begin{array}{l}
\mathbf{H}_i\Bigl |\Psi \Bigr >=H_i\Bigl |\Psi \Bigr >\, , \quad i=1, \ldots , L
\\ \mathbf{M}_1\Bigl |\Psi \Bigr >=M_1\Bigl |\Psi \Bigr >.
\end{array}\right. 
\eeq

Another distinguished set of commuting ``Hamiltonians'' is
\beq\label{6v12}
{\bf G}_i={\bf T}^{(h)}(x_i-\eta ).
\eeq
It can be shown that
\beq\label{6v13}
{\bf G}_i {\bf H}_i =\prod_{k\neq i}^L
\frac{\sinh (x_i \! -\! x_k \! + \! \eta )}{\sinh (x_i \! -\! x_k)}\, {\bf 1}.
\eeq

\paragraph{The Bethe ansatz solution.}
The operators $\mathbf{T}^{h,v}(x)$, $\mathbf{S}^z$ can be diagonalized simultaneously 
for any $x$. Below we will work with the operator 
$\mathbf{T}^{(h)}(x)$ or, equivalently, with the set of 
commuting ``Hamiltonians'' $\mathbf{H}_k$
(they generalize Hamiltonians of the trigonometric Gaudin model). 
This problem is usually solved
by the algebraic Bethe ansatz \cite{QISM1,QISM2}.
In the sector where $\mathbf{S}^z$ has eigenvalue $S^z=L-2M_2\geq 0$ 
the eigenvalues $T(x)$ of $\mathbf{T}^{(h)}(x)$ are given by the formula
\beq\label{ba1}
T(x)=e^{Lh}\! \prod_{k=1}^L \frac{\sinh (x\! -\! x_k\! +\! \eta )}{\sinh (x-x_k)}
\prod_{\alpha =1}^{M_2}\frac{\sinh (x\! -\! u_\alpha \! -\! \eta )}{\sinh (x-u_\alpha )}
+e^{-Lh}\! \prod_{\alpha =1}^{M_2}\frac{\sinh (x\! -\! 
u_\alpha \! +\! \eta )}{\sinh (x-u_\alpha )}
\eeq
(recall that $M_1+M_2=L$).
The Bethe roots $u_{\alpha}$ are to be found from the system of Bethe equations
\beq\label{ba2}
e^{2Lh}\prod_{k=1}^L \frac{\sinh (u_\alpha \! -\! x_k\! +\! \eta )}{\sinh (u_\alpha -x_k)}=
\prod_{\beta =1, \neq \alpha}^{M_2} 
\frac{\sinh (u_{\alpha}\! -\! u_\beta \! +\! 
\eta )}{\sinh (u_{\alpha}\! -\! u_\beta \! -\! \eta )}.
\eeq
The corresponding eigenvalues of $\mathbf{H}_j$ and $\mathbf{G}_j$ are
\beq\label{ba3}
H_j= e^{Lh}\! \prod_{k=1, \neq j}^L  \frac{\sinh (x_j\! -\! x_k\! +
\! \eta )}{\sinh (x_j-x_k)}
\prod_{\alpha =1}^{M_2}\frac{\sinh (x_j\! -\! u_\alpha \! -\! 
\eta )}{\sinh (x_j-u_\alpha )},
\eeq
\beq\label{ba4}
G_j= e^{-Lh}\! 
\prod_{\alpha =1}^{M_2}\frac{\sinh (x_j\! -\! u_\alpha )}{\sinh (x_j-u_\alpha \! -\! 
\eta )}.
\eeq

\section{The trigonometric Ruijsenaars-Schneider model}

The Ruijsenaars-Schneider (RS) system of particles \cite{Ruijs1} is the relativistic 
generalization of the Calogero-Moser-Sutherland model. Let $p_i$, $x_i$ 
be canonical variables with the Poisson brackets 
$\{p_i, x_j\}=\delta_{ij}$. The trigonometric (hyperbolic) RS system of $L$ 
particles is defined by the classical
Hamiltonian 
\beq\label{rs0}
{\cal H}=\sum_{i=1}^L e^{\eta p_i}\prod_{k\neq i}^L
\frac{\sinh (x_i-x_k + \eta )}{\sinh (x_i-x_k)},
\eeq
where the parameter $\eta$ has the meaning of the inverse velocity of light.
The velocities of particles are given by
\beq\label{rs4}
\dot x_i=\frac{\p {\cal H}}{\p p_i}=\eta
e^{\eta p_i}\prod_{k\neq i}^L
\frac{\sinh (x_i-x_k + \eta )}{\sinh (x_i-x_k)}.
\eeq
The equations of motion $\displaystyle{\dot p_i=-\frac{\p {\cal H}}{\p x_i}}$ are
\beq\label{rs1}
\ddot x_j= - \sum_{k=1, \neq j}^L 
\frac{2\dot x_j \, \dot x_k \, \sinh^2 \eta \, 
\cosh (x_j\! - \! x_k)}{\sinh (x_j\! - \! x_k\! +\! \eta )
\sinh (x_j\! - \! x_k)
\sinh (x_j\! - \! x_k\! -\! \eta )}.
\eeq

Two interesting 
special cases of the trigonometric RS model are $\eta =\pm \infty$ and 
$\eta =i\pi /2$. In the former case the equations of motion simplify to
\beq\label{rs2}
\eta =\pm \infty : \quad \ddot x_j = 2\! \sum_{k=1, \neq j}^L \!
\dot x_j  \dot x_k  \coth (x_j\! - \! x_k).
\eeq
In the latter case they have the form
\beq\label{rs3}
\eta =\frac{i\pi}{2}: \quad \ddot x_j = 4\! \sum_{k=1, \neq j}^L 
\frac{\dot x_j  \dot x_k }{\sinh 2(x_j\! - \! x_k)}.
\eeq

The RS model is known to be integrable. It has the Lax representation 
$
\dot \mathsf{L}=[\mathsf{A}, \, \mathsf{L}]
$
with the Lax matrix\footnote{The Lax matrix used in \cite{BLZZ16} is 
$\tilde \mathsf{L}=-\mathsf{L}^{\sf t}$, where ${\sf t}$ means 
transposition.}
\beq\label{lax1}
\mathsf{L}_{ij}=\mathsf{L}_{ij}(\{x_k\}_L, \, \{ \dot x_k \}_L)=
\frac{\sinh \eta \,\, \dot x_i}{\sinh (x_i\! - \! x_j\! -\! \eta )}
\eeq
and the ${\sf A}$-matrix
\beq\label{lax2}
\mathsf{A}_{jk}=\Bigl ( \sum_{l\neq j}\dot x_l \coth (x_j\! - \! x_l) -
\sum_l \dot x_l \coth (x_j\! - \! x_l\! +\! \eta )\Bigr ) \delta_{jk} +
\frac{1-\delta_{jk}}{\sinh (x_j\! - \! x_k)}
\eeq
The Lax representation implies that the time evolution of the Lax matrix 
is a similarity transformation: ${\sf L}(t)={\sf U}(t){\sf L}{\sf U}^{-1}(t)$.
In terms of momenta we have:
\beq\label{lax3}
\mathsf{L}_{ij}=
\frac{\eta \, \sinh \eta}{\sinh (x_i\! - \! x_j\! -\! \eta )}\, e^{\eta p_i}
\prod_{k\neq i}^L
\frac{\sinh (x_i-x_k + \eta )}{\sinh (x_i-x_k)}.
\eeq
Note that ${\cal H}=-\eta ^{-1}\mbox{tr}\, \mathsf{L}$. The  
integrals of motion in involution are given by
\beq\label{31}
{\cal H}_k=\mbox{tr}\, \mathsf{L}^k\,, \quad {\cal H}=-\eta ^{-1}{\cal H}_1.
\eeq 
The generating function of conserved quantities is the characteristic 
polynomial $Q(\lambda )=\det (\lambda {\sf I}-{\sf L})$.

Let $\mathsf{X}=\mathrm{diag} (x_1, x_2, \ldots , x_L)$ be the diagonal matrix 
with the diagonal entries being coordinates of the particles.
It is easy to check that the matrices $\mathsf{X}, \, \mathsf{L}$ satisfy the 
relation
$$
e^{-\eta}e^{\mathsf{X}}\mathsf{L}e^{-\mathsf{X}}-
e^{\eta}e^{-\mathsf{X}}\mathsf{L}e^{\mathsf{X}}=2 \; \sinh\eta \;  \dot \mathsf{X} \mathsf{E},
$$
where $\mathsf{E}$ is the $L\! \times \! L$ matrix of rank 1 with all entries equal to 1.

The Lax matrix of the RS model 
admits a simple factorization:
\beq\label{lax4}
\mathsf{L}=\dot \mathsf{X}\mathsf{C},
\eeq
where $\mathsf{X}=\mathrm{diag} (x_1, x_2, \ldots , x_L)$ and 
$\mathsf{C}$ is the trigonometric Cauchy matrix 
$$\displaystyle{\mathsf{C}_{ij}=\frac{\sinh \eta}{\sinh (x_i\! -\! x_j \! -\! \eta )}}.$$
It allows one to calculate the characteristic 
polynomial explicitly. We use the known fact that the coefficient 
in front of $\lambda ^{L-k}$ in the polynomial
$\det_{L\times L}(\lambda \mathsf{I} +\mathsf{M})$ equals 
the sum of all diagonal $k\! \times \! k$ 
minors of the matrix $\mathsf{M}$. All such minors 
can be found using decomposition (\ref{lax4})
and the explicit expression for the determinant
$$
\det_{1\leq i,j\leq n} \frac{\sinh \eta}{\sinh (x_i\! -\! x_j \! -\! \eta )}=(-1)^n \!
\prod_{1\leq i<j\leq n} C(x_i \! - \! x_j), \quad
C(x)=\frac{\sinh ^2 x}{\sinh (x \! +\! \eta ) \sinh (x \! -\! \eta )}
$$
Therefore,
\beq\label{lax51}
\det_{L\times L}(\lambda \mathsf{I}-\mathsf{L})=
\sum_{n=0}^L (-1)^n\mathcal{E}_n \lambda^{L-n},
\eeq
where
\beq\label{lax61}
{\cal E}_n = (-1)^n
\sum_{1\leq i_1<\ldots < i_n \leq L}
\dot x_{i_1}\ldots \dot x_{i_n}
\prod_{1\leq \alpha < \beta \leq n}
C(x_{i_\alpha} \! - \! x_{i_\beta}).
\eeq
The integrals of motion ${\cal E}_n$ are related to the 
integrals of motion ${\cal H}_n$ by the Newton's formula
$\sum_{k=0}^L(-1)^k{\cal E}_{L-k}{\cal H}_k=0$, where
${\cal H}_0=\mbox{tr}\, {\sf L}^0 =L$.

In fact the Lax matrix admits another factorizaton \cite{BLZZ16} which is non-trivial: 
\beq\label{lax5}
\mathsf{L}=- \eta \, e^{\eta \mathsf{P}}\, {\sf D}_{\eta}
(\mathsf{V}^{\sf t})^{-1}\, \mathsf{S}^{-1}\, \mathsf{V}^{\sf t}\, ({\sf D}_{\eta})^{-1}
\eeq
(see \cite{AASZ14,LOZ14,GZZ14} for a similar representation
in the rational case).
Here 
${\sf P}, {\sf D}, {\sf S}$ are diagonal matrices
$\mathsf{P}=\mathrm{diag} (p_1, p_2, \ldots , p_L)$,
\beq\label{lax6}
({\sf D}_{\eta})_{ij}=\delta_{ij}\prod_{k\neq i}^L \sinh (x_i-x_k +\eta )
\eeq
\beq\label{lax7}
{\sf S}_{ij}=\delta_{ij}e^{-(2i-L-1)\eta }
\eeq
and ${\sf V}$ is the Vandermonde type matrix
\beq\label{lax8}
{\sf V}_{ij}=e^{(2j-L-1)x_i }.
\eeq
Equation (\ref{lax5}) is the classical version of the factorized $L$-operator
for the quantum trigonometric RS model \cite{AHZ97}.

\section{The correspondence between the 6-ver\-tex model and the RS model}

Consider the Lax matrix (\ref{lax1}) of the $L$-particle RS model, where 
the coordinates of particles, $x_i$, are identified with the inhomogeneity parameters and the 
inverse ``velocity of light'', $\eta$, is identified with the anisotropy parameter.
Let us also substitute $\dot x_i = -H_i$ and consider the matrix
$\mathsf{L}=\mathsf{L}(\{ x_i\}_L, \{ - H_i\}_L)$:
{\small
\beq\label{cor1}
\mathsf{L}
%(\{ x_i\}, \{ -\sinh \eta \, H_i\})=
=\left (
\begin{array}{ccccc}
H_1 & \displaystyle{\frac{\sinh \eta \, H_1}{\sinh (x_2 \! -\! x_1\! +\! \eta )}} &
\displaystyle{\frac{\sinh \eta \, H_1}{\sinh (x_3 \! -\! x_1\! +\! \eta )}} & \ldots &
\displaystyle{\frac{\sinh \eta \, H_1}{\sinh (x_L \! -\! x_1\! +\! \eta )}} 
\\ &&&& \\
\displaystyle{\frac{\sinh \eta \, H_2}{\sinh (x_1 \! -\! x_2\! +\! \eta )}} & H_2 &
\displaystyle{\frac{\sinh \eta \, H_2}{\sinh (x_3 \! -\! x_2\! +\! \eta )}} & \ldots &
\displaystyle{\frac{\sinh \eta \, H_2}{\sinh (x_L \! -\! x_2\! +\! \eta )}} 
\\ &&&& \\
\vdots & \vdots & \vdots & \ddots & \vdots
\\ &&&& \\
\displaystyle{\frac{\sinh \eta \, H_L}{\sinh (x_1 \! -\! x_L\! +\! \eta )}} &
\displaystyle{\frac{\sinh \eta \, H_L}{\sinh (x_2 \! -\! x_L\! +\! \eta )}} &
\displaystyle{\frac{\sinh \eta \, H_L}{\sinh (x_3 \! -\! x_L\! +\! \eta )}} & \ldots &
H_L
\end{array}
\right ).
\eeq
}
The correspondence between the 6-ver\-tex model and the RS model
consists in the fact that if the $H_k$'s are eigenvalues of the 
operators ${\bf H}_k$, then the eigenvalues of the RS Lax matrix are
\beq\label{alg2}
\begin{array}{l}
e^{Lh -(M_1-1)\eta +2\eta j}, \quad j=0, 1, \ldots , M_1-1,
\\ \\
e^{-Lh-(M_2-1)\eta +2\eta j}, \quad j=0, 1, \ldots , M_2-1.
\end{array}
\eeq
In the terminology of the Bethe ansatz technique, they form 
``strings'' of lengths $M_1, M_2$ centered at $e^{\pm Lh}$.
We see that the spectrum of ${\sf L}$ depends only on the horizontal 
external field $h$ (and on $M_1, M_2$). This allows one to say that the spectral problem
for the 6-vertex transfer matrix is equivalent to the following 
{\it inverse} spectral problem for the RS Lax matrix: given $x_i$'s,
to find $H_i$'s in such a way that the eigenvalues of the Lax matrix have the 
fixed prescribed values (\ref{alg2}).
Equivalently, one fixes the values of the RS integrals of motion to be
\beq\label{int1}
{\cal H}_n =\mbox{tr}\, {\sf L}^n=
e^{Lhn}\, \frac{\sinh (M_1\eta n)}{\sinh (\eta n)}+
e^{-Lhn}\, \frac{\sinh (M_2n\eta )}{\sinh (n\eta )}.
\eeq

According to (\ref{lax51}), (\ref{lax61}), we have (\ref{lax51}),
where
\beq\label{QC5}
{\cal E}_n = 
\sum_{1\leq i_1<\ldots < i_n \leq L}
H_{i_1}\ldots H_{i_n}
\prod_{1\leq \alpha < \beta \leq n}
\frac{\sinh ^2(x_{i_\alpha} \! - \! 
x_{i_\beta})}{\sinh (x_{i_\alpha} \! - \! x_{i_\beta}\! +\! 
\eta )\sinh (x_{i_\alpha} \! - \! x_{i_\beta}\! -\! 
\eta )}.
\eeq
The spectrum of the operators ${\bf H}_i$ can be found by solving the 
algebraic equations
\beq\label{alg1}
{\cal E}_n =e_n, \quad n=1, \ldots , L,
\eeq
where $e_n$ are elementary symmetric functions,
$\displaystyle{
e_n = 
\sum_{1\leq i_1<\ldots < i_n \leq L}
\xi_{i_1}\ldots \xi_{i_n},}
$
of the variables $\xi_i$ which are taken from the set
$$
\begin{array}{lll}
\{\xi _k \}_L &=&
\Bigl \{\underbrace{e^{Lh-(M_1\! -\! 1)\eta }, e^{Lh-(M_1\! -\! 3)\eta }, \ldots 
e^{Lh+(M_1\! -\! 1)\eta }}_{M_1}, \,\,
\\ && \\
&& \hspace{2cm}
\underbrace{e^{-Lh-(M_2\! -\! 1)\eta }, e^{-Lh-(M_2\! -\! 3)\eta }, \ldots 
e^{-Lh+(M_2\! -\! 1)\eta }}_{M_2}
\Bigr \}
\end{array}
$$
of eigenvalues of the matrix ${\sf L}$.
These equations have many solutions which correspond to different
eigenstates.

It is interesting to note that eigenvalues of the commuting 
Hamiltonians ${\bf G}_i$ are related to {\it momenta} of the RS particles.
More precisely, it follows from (\ref{6v13}) and (\ref{rs4}) that 
as soon as we identify $H_i=-\dot x_i$ we should also identify
\beq\label{alg2a}
G_i=-\eta^{-1}e^{-\eta p_i}, \quad i=1, \ldots , L.
\eeq

\section{Proof of the correspondence}

The proof of the correspondence is straightforward but rather 
involved. In particular, it employs the non-trivial factorization (\ref{lax5})
of the Lax matrix.

First let us prove the following lemma \cite{BLZZ16}.

\noindent
{\bf Lemma 1.} 
{\it 
Let ${\sf Q}$, $\widetilde {\sf Q}$ be
a pair of $N\times N$ and $M\times M$ matrices
\beq\label{ap1}
{\sf Q}_{ij}\Bigl ( \{x_i\}_N, \{y_{\alpha}\}_M, g\Bigr )
=\displaystyle{ \frac{g\sinh \eta }{\sinh (x_j \! -\! x_i \! +\!\eta )}
\prod_{k\neq i}^N\frac{\sinh (x_i \! -\! x_k \! +\!\eta )}{\sinh (x_i \! -\! x_k)}
\prod_{\gamma =1}^M\frac{\sinh (x_i \! -\! y_{\gamma})}{\sinh (x_i \! -\! y_{\gamma}
\! +\! \eta )}
}
\eeq
where $i,j =1, \ldots , N$ and
\beq\label{ap2}
\widetilde {\sf Q}_{\alpha \beta}\Bigl ( \{y_{\gamma}\}_M, \{x_i\}_N,  g\Bigr )
=\displaystyle{ \frac{g\sinh \eta }{\sinh (y_\beta \! -\! y_\alpha \! +\!\eta )}
\prod_{\gamma\neq \alpha}^M
\frac{\sinh (y_\alpha \! -\! y_\gamma \! -\!\eta )}{\sinh (y_\alpha \! -\! y_\gamma)}
\prod_{k =1}^N\frac{\sinh (y_\alpha \! -\! x_k)}{\sinh (y_\alpha \! -\! x_k
\! -\! \eta )}
}
\eeq
where $\alpha , \beta =1, \ldots , M$ (for definiteness, we assume that 
$M\leq N$). Then the following identity holds true:
\beq\label{ap3}
\det_{N\times N}\Bigl (\lambda {\sf I}-{\sf Q}
\bigl ( \{x_i\}_N, \{y_{\alpha}\}_M, g\bigr )
\Bigr )
=\det_{(N\! -\! M)\times (N\! -\! M)}( \lambda {\sf I}-g{\sf S}_{N-M})
\det_{M\times M} \Bigl (\lambda {\sf I}-
\widetilde {\sf Q}\bigl ( \{y_{\alpha}\}_M, \{x_i\}_N,  g\bigr )
\Bigr )
\eeq
Here we use 
the notation $({\sf S}_{K})_{ij}=\delta_{ij}e^{-(2i-K-1)\eta}$ ($i,j=1, \ldots , K$)
for the matrix of the form (\ref{lax7}) of size $K\times K$.
}

\noindent
This means that the matrix ${\sf Q}$ (\ref{ap1}) has $N-M$ eigenvalues of the form
$ge^{-(2i-N+M-1)\eta}$, $i=1, \ldots, N-M$. In particular, at $M=0$ we have
\beq\label{ap4}
\det_{N\times N}\Bigl (\lambda {\sf I}-{\sf Q}
\bigl ( \{x_i\}_N, \emptyset , g\bigr )
\Bigr )
=\det_{N\times N}( \lambda {\sf I}-g{\sf S}_{N})=\prod_{i=0}^{N-1}
(\lambda -ge^{-(2i-N+1)\eta}).
\eeq

\noindent
{\it Proof.} The both sides of (\ref{ap3}) are rational functions of $t_i=e^{2x_i}$.
It is enough to prove that they have the same residues at the poles 
and the same values at infinity.

For the proof we need the factorization of the matrices 
${\sf Q}$, $\widetilde {\sf Q}$ which is similar to (\ref{lax5}):
\beq\label{ap5}
{\sf Q}\Bigl ( \{x_i\}_N, \{y_{\alpha}\}_M, g\Bigr )=
g{\sf W}^{(N,M)}{\sf D}_{\eta}(\{x_i\}_N)
({\sf V^t})^{-1}(\{x_i\}_N){\sf S}_{N}^{-1}{\sf V^t}(\{x_i\}_N)
{\sf D}_{\eta}^{-1}(\{x_i\}_N),
\eeq
\beq\label{ap6}
\widetilde {\sf Q}\Bigl ( \{y_\alpha\}_M, \{x_{i}\}_N, g\Bigr )=
g\widetilde {\sf W}^{(N,M)}{\sf D}_{0}^{-1}(\{y_\alpha\}_M)
{\sf V}(\{y_\alpha\}_M){\sf S}_{M}{\sf V}^{-1}(\{y_\alpha \}_M)
{\sf D}_{0}(\{y_\alpha \}_M).
\eeq
Here
\beq\label{ap7}
{\sf V}_{ij}(\{q_k\}_K)=e^{(2j-K-1)q_i}, \quad i,j=1, \ldots , K,
\eeq
\beq\label{ap8}
({\sf D}_\xi )_{ij}(\{q_k\}_K)= \delta_{ij}
\prod_{k\neq j}^K\sinh (q_i-q_k+\xi ), \quad i,j=1, \ldots , K,
\eeq
\beq\label{ap9}
{\sf W}^{(N,M)}_{ij}=\delta_{ij}\prod_{\gamma =1}^M
\frac{\sinh (y_\gamma -x_i)}{\sinh (y_\gamma -x_i -\eta )}\,, \quad
i,j=1, \ldots , N,
\eeq
\beq\label{ap10}
\widetilde {\sf W}^{(N,M)}_{\alpha \beta}=\delta_{\alpha \beta}\prod_{k =1}^N
\frac{\sinh (y_\alpha -x_k)}{\sinh (y_\alpha -x_k -\eta )}\,, \quad
\alpha , \beta =1, \ldots M. 
\eeq
Let us note that $\det {\sf W}^{(N,M)}=\det \widetilde {\sf W}^{(N,M)}$.
Hence the statement of the lemma acquires the form
\beq\label{ap3a}
\begin{array}{c}
\displaystyle{
\det_{N\times N}\Bigl (\lambda ({\sf W}^{N,M})^{-1}-{\sf Q}_0
\bigl ( \{x_i\}_N, g\bigr )
\Bigr )}\\ \\ \displaystyle{
=\det_{(N\! -\! M)\times (N\! -\! M)}( \lambda {\sf I}-g{\sf S}_{N-M})
\det_{M\times M} \Bigl (\lambda (\widetilde {\sf W}^{N,M})^{-1}-
\widetilde {\sf Q}_0\bigl ( \{y_{\alpha}\}_M, g\bigr )
\Bigr ),}\end{array}
\eeq
where ${\sf Q}_0
\bigl ( \{x_i\}_N, g\bigr )={\sf Q}
\bigl ( \{x_i\}_N, \emptyset , g\bigr )$, $\widetilde {\sf Q}_0
\bigl ( \{y_\alpha\}_M, g\bigr )=\widetilde {\sf Q}
\bigl ( \{y_\alpha\}_M, \emptyset , g\bigr )$.

First let us prove that the left hand side of (\ref{ap3a}) does not have 
poles at the points $x_i=x_k$ and $x_i=x_k+\eta$. For this we write
$$
{\sf V}_{ij}(\{x_k\}_N)=e^{(1-N)x_i}\Bigl (e^{2x_i}\Bigr )^{j-1}
={\sf T}_{ii}(\{x_k\}_N)\widetilde {\sf V}_{ij}(\{x_k\}_N),
$$
where $\widetilde {\sf V}$ is the Vandermonde matrix of variables
$t_i=e^{2x_i}$ ($\widetilde {\sf V}_{ij}=t_i^{j-1}$) and 
${\sf T}$ is the diagonal matrix 
(${\sf T}_{ii}=e^{(1-N)x_i}$).
Then one can rewrite
the left hand side of (\ref{ap3a}) as
$$
\begin{array}{ll}
&\displaystyle{\det_{N\times N}\Bigl (\lambda ({\sf W}^{N,M})^{-1}-{\sf Q}_0
\bigl ( \{x_i\}_N, g\bigr )
\Bigr )}
\\ &\\
=& \displaystyle{\det_{N\times N}\Bigl (\lambda ({\sf W}^{N,M})^{-1}-
g{\sf D}_{\eta}({\sf V^t})^{-1}{\sf S}_{N}^{-1}\, {\sf V^t}\, {\sf D}_{\eta}^{-1}
\Bigr )}
\\ &\\
=& \displaystyle{\det_{N\times N}\Bigl (\lambda ({\sf W}^{N,M})^{-1}-
g{\sf T}^{-1}(\widetilde {\sf V}^{\sf t})^{-1}{\sf S}_{N}^{-1}\, 
\widetilde {\sf V}^{\sf t}\, {\sf T}
\Bigr )}
\\ &\\
=& \displaystyle{
\det_{N\times N}\Bigl (\lambda \widetilde {\sf V}^{\sf t}({\sf W}^{N,M})^{-1}
(\widetilde {\sf V}^{\sf t})^{-1}-g {\sf S}_N^{-1}
\Bigr ).}
\end{array}
$$
The inverse to the Vandermonde matrix is given by the explicit 
expression
$$
\left. (\widetilde {\sf V}^{\sf t})^{-1}_{kj}=\frac{1}{(j-1)!}\, 
\p_s^{j-1}\prod_{l\neq k}^N\frac{s-t_l}{t_k-t_l}\,
\right |_{s=0}.
$$
Then the matrix element
$\Bigl (\widetilde {\sf V}^{\sf t}{\sf W}^{-1}
(\widetilde {\sf V}^{\sf t})^{-1}\Bigr )_{ij}$
has the form
$$
\Bigl (\widetilde {\sf V}^{\sf t}{\sf W}^{-1}
(\widetilde {\sf V}^{\sf t})^{-1}\Bigr )_{ij}=
\sum_{k=1}^N \widetilde {\sf V}_{ki}\, {\sf W}^{-1}_{kk}\,
(\widetilde {\sf V}^{\sf t})^{-1}_{kj}
$$
$$
\left. =\,\sum_{k=1}^N t_k^{i-1}\, {\sf W}^{-1}_{kk}\,
\frac{1}{(j-1)!}\, 
\p_s^{j-1}\prod_{l\neq k}^N\frac{s-t_l}{t_k-t_l}\,
\right |_{s=0}.
$$
The expression
$$
\sum_{k=1}^N t_k^{i-1}\, {\sf W}^{-1}_{kk}\,
\prod_{l\neq k}^N\frac{s-t_l}{t_k-t_l}\, =\sum_{k=1}^N
\Bigl (\widetilde {\sf V}^{\sf t}{\sf W}^{-1}
(\widetilde {\sf V}^{\sf t})^{-1}\Bigr )_{ik}s^k
$$
is the generating function of the matrix elements.
We see that the poles at $x_a=x_b+\eta$ are absent. The pole at
$x_a=x_b$ comes from the terms with $k=a,b$.
The residue at this point is given by the expression
$$
\prod_{m=1}^N(s-t_m)\left (
\frac{t_a^{i-1}}{s-t_a}\,
\frac{{\sf W}^{-1}_{aa}}{\prod_{l\neq a,b}(t_a-t_l)}-
\frac{t_b^{i-1}}{s-t_b}\,
\frac{{\sf W}^{-1}_{bb}}{\prod_{l\neq a,b}(t_b-t_l)}
\right )
$$
which is zero at $x_a=x_b$.

In a similar way, one can show that there are no poles at
$y_\alpha =y_\beta$ and $y_\alpha =y_\beta +\eta$ in the right hand side
of (\ref{ap3a}). This means that the both sides have poles only 
at the points $x_i=y_\alpha$.

The next step is induction in $M$. At $M=0$ we have
$$
\begin{array}{c}
\displaystyle{
\det_{N\times N} (\lambda {\sf I}-{\sf Q})=\det_{N\times N} \Bigl (\lambda {\sf I}-
g{\sf D}_{\eta}({\sf V^t})^{-1}{\sf S}_{N}^{-1}\, {\sf V^t}\, {\sf D}_{\eta}^{-1}
\Bigr )
}
\\ \\
\displaystyle{
=\, \det_{N\times N} (\lambda {\sf I}-g{\sf S}_N^{-1})=
\det_{N\times N} (\lambda {\sf I}-g{\sf S}_N)
}
\end{array}
$$
which agrees with the statement of the lemma (the second
determinant of the $0\times 0$ matrix in (\ref{ap3}) is set to be equal to 1).
The assumption of the induction is that the statement of the lemma 
holds true at $M-1$ and for any $N\geq M-1$. 
Pass from $M-1$ to $M$ and consider the residue at $x_i=y_\alpha$
in the left hand side of (\ref{ap3a}):
$$
\mbox{res}_{x_i=y_{\alpha}}
\det_{N\times N}\Bigl (\lambda ({\sf W}^{N,M})^{-1}-
{\sf Q}_0(\{x_k\}_N, g)
\Bigr )
$$
$$
=\det_{(N-1)\times (N-1)}\Bigl (\lambda ({\sf W}^{N-1,M})^{-1}-
{\sf Q}_0^{ii}(\{x_k\}_N, g)
\Bigr )\, \times \, 
\lambda \sinh \eta \prod_{\gamma \neq \alpha}^M 
\frac{\sinh (x_i\! -\! y_{\gamma}\! +\! \eta )}{\sinh (x_i\! -\! y_{\gamma})}
$$
$$
=\det_{(N-1)\times (N-1)}\Bigl (\lambda ({\sf W}^{N-1,M-1})^{-1}-
{\sf Q}_0(\{x_k\}_N\setminus x_i, g)
\Bigr )\, \times \, 
\lambda \sinh \eta 
$$
$$
\times \prod_{k=1, \neq i}^N 
\frac{\sinh (x_k\! -\! y_{\alpha}\! +\! \eta )}{\sinh (x_k\! -\! y_{\alpha})}
\prod_{\gamma =1, \neq \alpha}^M
\frac{\sinh (x_i\! -\! y_{\gamma}\! +\! \eta )}{\sinh (x_i\! -\! y_{\gamma})}.
$$
In the second line ${\sf Q}_0^{ii}$ is the matrix ${\sf Q}_0$ without its $i$-th row and
$i$-th column. In a similar way, the residue in the right hand side of (\ref{ap3a}) is
$$
\mbox{res}_{x_i=y_{\alpha}}
\det_{M\times M}\Bigl (\lambda (\widetilde {\sf W}^{N,M})^{-1}-
\widetilde{\sf Q}_0(\{y_\gamma\}_M, g)
\Bigr )
$$
$$
=\det_{(M-1)\times (M-1)}\Bigl (\lambda (\widetilde{\sf W}^{N,M-1})^{-1}-
\widetilde{\sf Q}_0^{\alpha \alpha}(\{y_\gamma\}_M, g)
\Bigr )\, \times \, 
\lambda \sinh \eta \prod_{k=1 \neq i}^N
\frac{\sinh (x_k\! -\! y_{\alpha}\! +\! \eta )}{\sinh (x_k\! -\! y_{\alpha})}
$$
$$
=\det_{(M-1)\times (M-1)}\Bigl (\lambda (\widetilde{\sf W}^{N-1,M-1})^{-1}-
\widetilde{\sf Q}_0(\{y_\gamma \}_M \setminus y_{\alpha}, g)
\Bigr )\, \times \, 
\lambda \sinh \eta 
$$
$$
\prod_{\gamma =1, \neq \alpha}^M
\frac{\sinh (x_i\! -\! y_{\gamma}\! +\! \eta )}{\sinh (x_i\! -\! y_{\gamma})}
\times \prod_{k=1, \neq i}^N 
\frac{\sinh (x_k\! -\! y_{\alpha}\! +\! \eta )}{\sinh (x_k\! -\! y_{\alpha})}.
$$
We see that the multipliers near the determinants in the both sides 
are the same, so the equality of the residues in (\ref{ap3a}) for $N,M$ is reduced to
(\ref{ap3a}) for $N-1, M-1$ which holds true according to the assumption
of the induction. Therefore, the poles and the residues in all variables 
in the both sides
of (\ref{ap3a}) are the same.

We have thus proved that
\beq\label{ap3b}
\begin{array}{c}
\displaystyle{
\det_{N\times N}\Bigl (\lambda ({\sf W}^{N,M})^{-1}-{\sf Q}_0
\bigl ( \{x_i\}_N, g\bigr )
\Bigr )}\\ \\ \displaystyle{
=\det_{(N\! -\! M)\times (N\! -\! M)}( \lambda {\sf I}-g{\sf S}_{N-M})
\det_{M\times M} \Bigl (\lambda (\widetilde {\sf W}^{N,M})^{-1}-
\widetilde {\sf Q}_0\bigl ( \{y_{\alpha}\}_M, g\bigr )
\Bigr ) +C_{N,M},}\end{array}
\eeq
where $C_{N,M}$ are some constants. They can be found from the limit
$y_{\alpha}\to \infty$. We have:
%$$
%\mbox{lim}_{x_i\to \infty}
%\det_{N\times N}\Bigl (\lambda ({\sf W}^{N,M})^{-1}-
%{\sf Q}_0(\{x_k\}_N, g)
%\Bigr )
%$$
%$$
%=\, \left (\lambda e^{M\eta}-ge^{(N-1)\eta}\right )
%\det_{(N-1)\times (N-1)}\Bigl (\lambda ({\sf W}^{N-1,M})^{-1}-
%{\sf Q}_0(\{x_k\}_N\setminus x_i, g)e^{-\eta}
%\Bigr ),
%$$
%$$
%\mbox{lim}_{x_i\to \infty}
%\det_{M\times M}\Bigl (\lambda (\widetilde {\sf W}^{N,M})^{-1}-
%\widetilde{\sf Q}_0(\{y_\gamma\}_M, g)
%\Bigr )=
%\det_{M\times M}\Bigl (\lambda e^{\eta}(\widetilde {\sf W}^{N-1,M})^{-1}-
%\widetilde{\sf Q}_0(\{y_\gamma\}_M, g)
%\Bigr ).
%$$
$$
\mbox{lim}_{y_\alpha\to \infty}
\det_{N\times N}\Bigl (\lambda ({\sf W}^{N,M})^{-1}-
{\sf Q}_0(\{x_k\}_N, g)
\Bigr )
=
\det_{N\times N}\Bigl (\lambda e^{-\eta}({\sf W}^{N,M-1})^{-1}-
{\sf Q}_0(\{x_k\}_N, g)
\Bigr ),
$$
$$
\mbox{lim}_{y_\alpha\to \infty}
\det_{M\times M}\Bigl (\lambda (\widetilde {\sf W}^{N,M})^{-1}-
\widetilde{\sf Q}_0(\{y_\gamma\}_M, g)
\Bigr )
$$
$$
=\, \left (\lambda e^{-N\eta}-ge^{-(M-1)\eta}\right )
\det_{(M-1)\times (M-1)}\Bigl (\lambda (\widetilde{\sf W}^{N,M-1})^{-1}-
\widetilde{\sf Q}_0(\{y_\gamma\}_N\setminus y_\alpha, g)e^{\eta}
\Bigr ).
$$
Using the trivially checked identity
$$
\det_{(N\! -\! M\! +\! 1)\times (N\! -\! M\! +\! 1)}
\left (\lambda e^{-\eta}{\sf I}-g{\sf S}_{N-M+1}\right )=
\left (\lambda e^{(M-N-1)\eta}-g\right )
\det_{(N\! -\! M)\times (N\! -\! M)}
\left (\lambda {\sf I}-g{\sf S}_{N-M}\right ),
$$
one can see that $C_{N,M}=C_{N, M-1}$.
But we know that $C_{N,0}=0$ for any $N$. Therefore, $C_{N,M}=0$ for any $N,M$ and the
lemma is proved.

\noindent
{\bf Theorem 1.} 
{\it Let $H_i$ be eigenvalues of the operators ${\bf H}_i$, then
spectrum of the matrix ${\sf L}\Bigl (\{x_k\}_L, \{\dot x_k=-H_k\}_L \Bigr )$
is the following:
$$
\mbox{{\rm Spec}}\, {\sf L}\Bigl (\{x_k\}_L, \{\dot x_k=-H_k\}_L \Bigr )
$$
$$
=\Bigl \{\underbrace{e^{Lh-(M_1\! -\! 1)\eta }, e^{Lh-(M_1\! -\! 3)\eta }, \ldots 
e^{Lh+(M_1\! -\! 1)\eta }}_{M_1}, \,\,
\underbrace{e^{-Lh-(M_2\! -\! 1)\eta }, e^{-Lh-(M_2\! -\! 3)\eta }, \ldots 
e^{-Lh+(M_2\! -\! 1)\eta }}_{M_2}
\Bigr \}.
$$
}

\noindent
Obviously, the statement of the theorem is equivalent to
\beq\label{pr0}
\det_{L\times L} \Bigl [ \lambda {\sf I}-{\sf L}
\Bigl (\{x_k\}_L, \{-H_k\}_L \Bigr )\Bigr ]=
\det_{M_1\times M_1} \Bigl [ \lambda {\sf I}-e^{Lh}{\sf S}_{M_1}\Bigr ] 
\det_{M_2\times M_2} \Bigl [ \lambda {\sf I}-e^{-Lh}{\sf S}_{M_2}\Bigr ].
\eeq

\noindent
{\it Proof.} 
The value of $H_k$ through the Bethe roots is given by equation (\ref{ba3}).
Substituting (\ref{ba3}) into ${\sf L}\Bigl (\{x_k\}_L, \{\dot x_k=-H_k\}_L \Bigr )$,
we see that
$$
{\sf L}\Bigl (\{x_k\}_L, \{\dot x_k=-H_k\}_L \Bigr )=
{\sf Q}(\{x_k-\eta \}_L, \{u_{\alpha}\}_{M_2},e^{Lh}\}
$$
where ${\sf Q}$ is given by (\ref{ap1}).
Lemma 1 implies that
\beq\label{pr1}
\det_{L\times L}\Bigl (\lambda {\sf I}-{\sf L}
\Bigr )
=\! \det_{(L\! -\! M_2)\times (L\! -\! M_2)}
\Bigl ( \lambda {\sf I}-e^{Lh}{\sf S}_{L-M_2}\Bigr )
\det_{M_2\times M_2} \Bigl (\lambda {\sf I}-
\widetilde {\sf Q}\bigl ( \{u_{\alpha}\}_{M_2}, \{x_i\! -\! \eta \}_L,  e^{Lh}\bigr )
\Bigr )
\eeq
with
$$
\widetilde {\sf Q}_{\alpha \beta}
\bigl ( \{u_{\alpha}\}_{M_2}, \{x_i\!-\! \eta \}_L,  e^{Lh}\bigr )=
\frac{e^{Lh}\sinh \eta }{\sinh (u_\beta \! -\! u_\alpha \! +\!\eta )}
\prod_{\gamma\neq \alpha}^{M_2}
\frac{\sinh (u_\alpha \! -\! u_\gamma \! -\!\eta )}{\sinh (u_\alpha \! -\! u_\gamma)}
\prod_{k =1}^L\frac{\sinh (u_\alpha \! -\! x_k\! +\! \eta )}{\sinh (u_\alpha \! -\! x_k)}
$$
Imposing the Bethe equations (\ref{ba2}), we have:
$$
\widetilde {\sf Q}_{\alpha \beta}
\bigl ( \{u_{\alpha}\}_{M_2}, \{x_i\!-\! \eta \}_L,  e^{Lh}\bigr )\Bigl |_{BE}=
\frac{e^{-Lh}\sinh \eta }{\sinh (u_\beta \! -\! u_\alpha \! +\!\eta )}
\prod_{\gamma\neq \alpha}^{M_2}
\frac{\sinh (u_\alpha \! -\! u_\gamma \! +\!\eta )}{\sinh (u_\alpha \! -\! u_\gamma)}
$$
$$
=\, {\sf Q}_{\alpha \beta}
\bigl ( \{u_{\alpha}\}_{M_2}, \emptyset , e^{-Lh}\bigr )
$$
The second determinant in (\ref{pr1}) is then equal to
$$
\det_{M_2\times M_2} \Bigl (\lambda {\sf I}-
\widetilde {\sf Q}\bigl ( \{u_{\alpha}\}_{M_2}, \{x_i\! -\! \eta \}_L,  e^{Lh}\bigr )
\Bigr )
=\det_{M_2\times M_2} \Bigl (\lambda {\sf I}-
{\sf Q}\bigl ( \{u_{\alpha}\}_{M_2}, \emptyset ,  e^{-Lh}\bigr )
\Bigr )
$$
$$
=\,\, \det_{M_2\times M_2} \Bigl ( \lambda {\sf I}-e^{-Lh}{\sf S}_{M_2}\Bigr ).
$$
(The second equality again follows from Lemma 1.)
Combining this with (\ref{pr1}), we get (\ref{pr0}).

\section{Conclusion}

In this paper we have discussed the extension of the
quantum-classical correspondence to the simplest model of the $XXZ$ type. The main result is that the methods
suggested in \cite{GZZ14} for models parametrized by rational 
functions are successfully applicable to this case as well. The
anisotropy parameter of the 6-vertex model determines the splitting of spectrum of
the related classical Lax matrix.

The known examples suggest that the quantum-classical correspondence is specific for integrable models.
However, its origin and range of generality still remain obscure.
One of the open questions is the extension to models parametrized by elliptic functions
(in particular, to the 8-vertex model).

At the same time, similar phenomena are encountered in somewhat different contexts. For
example, the interpretation of the Planck constant as relativistic
deformation parameter is used in construction of the relativistic
Euler-Arnold tops \cite{LOZ14} via quasi-classical description of the
one-site spin chain. In some particular cases these tops are gauge
equivalent to the Ruijsenaars-Schneider models, and the Planck
constant entering the quantum $R$-matrix is identified with the relativistic parameter.
Another example is the Matsuo-Cherednik approach \cite{Matsuo,Cherednik2} to the quantum
Calogero-Moser models. The spectral parameters of the classical
$r$-matrices in the Knizhnik-Zamolodchikov connections play the role of
particle's coordinates in the Calogero-Moser models. Close topics
are discussed in \cite{MTV12}. At last, let us point out one more
interrelation between classical and quantum integrable systems
similar to the quantum-classical correspondence. It was observed in
\cite{NRS11} that the eigenstates of quantum Hitchin system
correspond to intersection points of two Lagrangian submanifolds in
the classical phase space given by the moduli of flat connections.

\section*{Acknowledgments}
This work has been funded by the  Russian Academic Excellence Project `5-100'.
The work of A.Liashyk has been also funded by joint NASU-CNRS project F14-2016.
The work of A.Zabrodin has been also supported in part by  
RFBR grant 14-02-00627.
The work of A.Zotov has been supported in part by
RFBR grants 14-01-00860 and 15-51-52031 HHC$_a$.
The authors thank the organizers of the workshop on classical and quantum
integrable systems (CQIS-2016) in St.-Petersburg, Russia, where some of the results 
were reported.

\end{document}